# SCENARIO BASED WORM TRACE PATTERN IDENTIFICATION TECHNIQUE

Siti Rahayu S., Robiah Y., Shahrin S., Mohd Zaki M., Irda R., Faizal M. A.
Faculty of Information and Communication Technology
Univeristi Teknikal Malaysia Melaka,
Durian Tunggal, Melaka,
Malaysia
.

*Abstract*—The number of malware variants is growing tremendously and the study of malware attacks on the Internet is still a demanding research domain. In this research, various logs from different OSI layer are explore to identify the traces leave on the attacker and victim logs, and the attack worm trace pattern are establish in order to reveal true attacker or victim. For the purpose of this paper, it will only concentrate on cybercrime that caused by malware network intrusion and used the traditional worm namely blaster worm variants. This research creates the concept of trace pattern by fusing the attacker's and victim's perspective. Therefore, the objective of this paper is to propose on attacker's, victim's and multi-step (attacker/victim)'s trace patterns by combining both perspectives. These three proposed worm trace patterns can be extended into research areas in alert correlation and computer forensic investigation.

## I. INTRODUCTION

Nowadays the numbers of cases of internet threat causes by malware have been tremendously increased. Malware that consist of Trojan, virus and worm had threatened the internet user and causes billion of losses to the internet users around the world. The exponential growth of malware variants as reported by AV-test.org [1] is quite alarming. This can be seen from the higher growth rate of the malware collection shown in the graph of the new malware samples over the last several years as depicted in Fig.1.

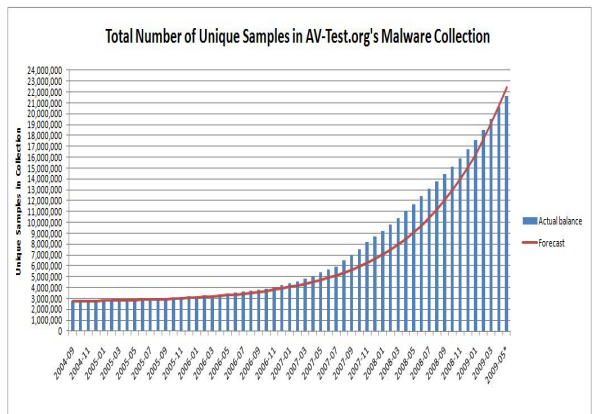

Fig. 1  The continuous growth of the number of malware [1]

In the report, the numbers of malware is not cumulative where they represent only the new variants in the time frame specified without including the previously identified ones.

This does not mean that there are only unique pieces of malware as there are also many variants of the same pieces of malware. Variants are often created to defeat the security tools, for instance a worm can mutate to a different variants, sometimes in only one hour [2]. Thus make it difficult for security tool to detect the threat.

Due to this reason the study of attacks on the internet has becoming an important research domain. The study on internet attack is very crucial, especially in developing an effective security tool to defend the internet user from the attack threat. Collecting data related to Internet threats has become a relatively common task for security researchers and from the data the researcher can investigate the attack pattern in order to find the root cause and effect of an attack. Attack pattern can also be used as a guide to the investigator for collecting and tracing the evidence in forensic field [3]. According to [4], the anatomy of attack consists of attacker and victim perspective.

To address this challenge, this paper propose on attacker's, victim's and multi-step (attacker/victim)'s trace patterns by collaborate both attacker and victim perspectives. This research explore the various logs from different OSI layer to identify the traces leave on the attacker and victim logs, and establish the attack trace pattern in order to reveal true attacker or victim. The research only focuses on cybercrime that caused by malware network intrusion specifically traditional worm namely blaster worm variants.

## II. RELATED WORK

### A. Blaster Worm

Generally malware consists of virus, worm and trojan horse [5]. For the purpose of this research, the researchers focuses on one types of worm that described by [6], which is traditional worm specifically blaster worm.

Blaster uses a public exploit for the RPC DCOM vulnerability in order to obtain a system-level command shell on its victims [7]. The worm start searching for IP addresses at a time for hosts with listening TCP port 135 open by sending a connection attempt to each one simultaneously. Once found, it tries to figure out the windows version and then sends the RPC DCOM exploit that binds a system level shell to TCP port 4444. Once the target is successfully compromised, the worm transmits the msblast.exe executable via TFTP on UDP port 69 to infect the host where the payloads used in the public DCOM exploit, as well as the TFTP functionality, are both encapsulated within msblast.exe. Once the executable has been transferred, or after 20 seconds have elapsed, the TFTP server is shut down and the worm then issues further commands to the victim to execute msblast.exe [8]. Assuming the executable was downloaded successfully, the propagation cycle then begins again from the newly infected host, while the infecting instance of the worm continues iterating through IP addresses.

### B. Trace Pattern

Trace is described as a process of finding or discovering the origin or cause of certain scenario and pattern is defined as a regular way in which certain scenario happened [9]. Trace pattern is essential in assisting the investigators tracing the evidence found at crime scenes. In the computer crime perspective, it can be found in any digital devices. These traces consist in a variety of data records their activities such as login and logout to the system, visit of pages, documents accesses, items created and affiliation to groups. Traces data are typically stored in log files and normally takes on several selected attributes such as port, action, protocol, source IP address and destination IP address.

The trace data can be used to identify a victim or attacker by analyzing the attack pattern which is represented in the form of trace pattern can helps determine how a crime is being committed. Attack pattern is type of pattern that is specified from attacker perspective. The pattern describes how an attack is performed, enumerates the security patterns that can be applied to defeat the attack, and describes how to trace the attack once it has occurred [10].

An attack pattern provides a systematic description of the attack goals and attack strategies for defending against and tracing the attack. Hence, attack patterns can guide forensic investigators in searching the evidence and the patterns can serve as a structured method for obtaining and representing relevant network forensic information. This also helps the forensic investigator at the data collection phase that requires the investigator to determine and identifying all the components to be collected, deciding the priority of the data, finding the location of the components and collecting data from each of the component during the investigation process [11].

Various descriptions provided by several researchers to describe the term attack pattern. According to [12], they use the term attack pattern to describe the steps in a generic attack. Meanwhile, [13] describe the term attack pattern as the attack steps, attack goal, pre-conditions and post-conditions of an attack. [14] describe an attack pattern as the approach used by attackers to generate an exploit against software in which it is a general framework for carrying out a particular type of attack, such as a method for exploiting a buffer overflow or an interposition attack that leverages certain kinds of architectural weaknesses.

Although there are different descriptions provided by several researchers, they have the same idea and concept that the attack pattern is very important to provide a way to protect them from any potential attack. For example, software developers use attack pattern to learn about how their software may be attacked. Armed with knowledge about potential attacks, developers can take steps to mitigate the impact of these attacks. Similarly, network administrator or network engineers use attack pattern to study on how the potential attacker attack their network in order to detect and block any vulnerabilities in their network.

The study from [12][13][14] discussed the concept of attack patterns as a mechanism to capture and communicate at the attacker's perspective that shows the common methods for exploiting software, system or network while [10] and [11] discussed on the attack pattern on how the attack is performed, the attack goals, how to defences against the attack and how to trace once it has occurred.

All of the researchers are only focusing on the attacker's perspective while victim's perspective is omitted. Therefore, this research proposed the trace patterns by focusing on the attacker's, victim's and attacker/victim's (multi-step) perspectives to provide clear view on how the attacker performed the attack and what is the impact caused by the attack.

Multi-step perspective proposed in this research is motivated based on the study by [15] in order to reveal the true attacker or victim. A multi-step attack is a sequence of attack steps that an attacker has performed, where each step of the attack is dependent on the successful completion of the previous step. These attack steps are the scan followed by the break-in to the host and the tool-installation, and finally an internal scan originating from the compromised host [16].

In the next section, researchers present the experimental approach used in this research to gather and analyse logs for designing the proposed worm trace pattern.

### III. EXPERIMENTAL APPROACH

This experimental approach used four phases: Network Environment Setup, Attack Activation, Trace Pattern Log Collection and Trace Pattern Log Analysis. Further details on these phases are explained as follows and depicted in Fig. 2.

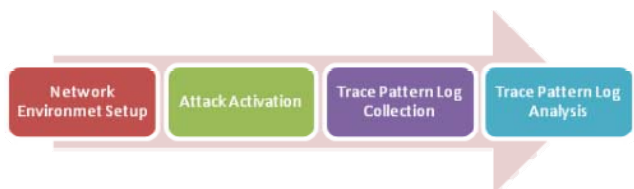

Fig. 2 Trace Pattern Experimental Approach Framework

### A. Network Environment Setup

The network environment setup use in this experiment is following the network simulation setup done by the MIT Lincoln Lab [17] and it only consists of Centos and Windows XP operating systems to suit our experiment's environment. The network design is shown in Fig. 3.

This network design consists of two switches, one router, three servers for Intrusion Detection System (IDS Arowana and IDS Sepat) and Network Time Protocol (NTP) run on *Centos 4.0*, seven victims and one attacker run on *Windows XP*.

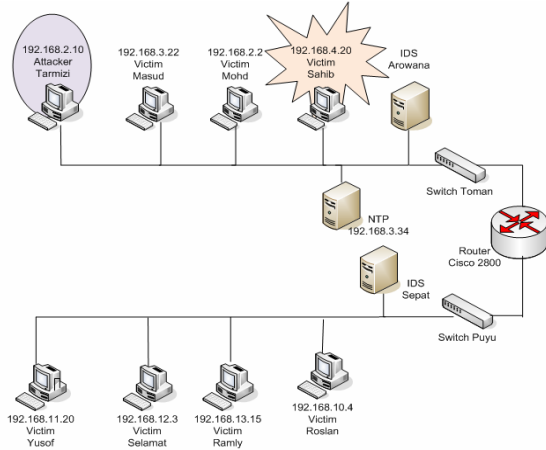

Fig. 3 Network Setup Environment for Blaster Trace Pattern

In this experiment, host 192.168.2.10 (*Tarmizi*) is Attacker, host 192.168.4.20 (*Sahib*) and 192.168.11.20 (*Yusof*) are the selected Victims for this research. The log files that are expected to be analyzed are *personal firewall log*, *security log*, *system log* and *application log*. Each log are generated by host level device and one log files by network level device (*alert log* by IDS). *Wireshark* were installed in each host and *tcpdump* script is activated in IDS to capture the whole network traffic. The *Wireshark* and *tcpdump* script were used to verify the traffic between particular host and other device.

### B. Attack Activation

The attacker machine, *Tarmizi* is selected to launch the Blaster variant attack. This experiment runs for one hour without any human interruption in order to obtain the attacker and victim logs.

### C. Trace Pattern Log Collection

Each machine generated *personal firewall log, security log, application log, system log* and *wireshark log*. The IDS machine generates *alert log* and *tcpdump log*. The trace pattern logs are collected at each victim and attacker machine. The *wireshark* and *tcpdump* files are used to verify the traffic between particular host and other device.

### D. Trace Pattern Log Analysis

In this trace pattern log analysis process the researchers analyze the logs generated by the attacker and victim machine. The objective of the trace pattern log analysis is to identify the victim and attacker trace pattern by observing the traces leave on the selected logs. The output from the analysis is used to construct worm trace pattern as proposed in Section V.

## IV. ANALYSIS AND FINDINGS

The researchers have collected all logs generated during the experiment and the attack scenario is identified. Based on the scenario, various logs from attacker's host, victim's host, and network are analyzed.

### A. Attack Scenario

The attack scenario as depicted in Fig. 4 is derived based on the log analysis in the experimental approach framework in section III.

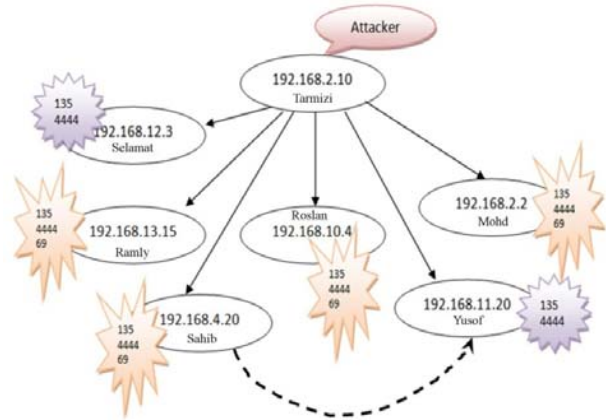

Fig. 4 Blaster Attack Scenario

Based on the analysis, the worm attack is activated in *Tarmizi* and successfully exploited all hosts that mark with *135, 4444* and *69* except for host *Yusof* and *Selamat*. These hosts was mark with *135* and *4444* and indicated that the attacker is already open the backdoor but unable to transfer the malicious codes through port *69*. Then, one of the infected hosts that are *Sahib* has organized attack on host *Yusof*.

### B. Trace Pattern Analysis

The purpose of trace pattern analysis is to construct the victim's, attacker's and multi-step (victim/attacker)'s trace pattern by observing the traces leave on the selected logs. The various logs involve in this analysis are host logs: *personal firewall log, security log, system log, application log* and network logs: *alert log* by IDS. Based on the attack scenario described previously, the logs are further analyzed to extract the trace pattern generated by the attack from the attacker's and victim's machine by tracing the fingerprints emerged in the logs.

In this analysis, the researchers have selected *Sahib* and *Yusof* as victims; and *Tarmizi* as attacker as shown in Fig. 3. The tracing tasks involve the log from devices at host and network level and initially started at victim's logs followed by attacker's logs: *personal firewall log, security log, system log* and *application log*. The network logs are used to complement the finding from the host level tracing tasks. The details of the analysis for victim, attacker and multi-step attack (attacker/victim) trace pattern are further elaborated in the next sub-section.

*1) Victim's Traces*

The victim's data traces are extracted from the logs at the victim's host and network log. The summary of the data traces are depicted in Fig. 5 and the evidences are found in *personal firewall log, security log, system log* and *application log* and the details of the traces are discussed.

| Level | Evidence/Log | Category | Trace Pattern | Extracted Data | Conceptual Diagram |
|---|---|---|---|---|---|
| Host | Personal Firewall Log | Scan and Exploit | **Action, Protocol and Destination Port:**<br>OPEN-INBOUND 135 TCP<br>OPEN-INBOUND 4444 TCP<br>OPEN 69 UDP<br><br>**Source IP Address and Destination IP Address:**<br>Refer to Table 1 | 2009-09-07 14:41:09 OPEN-INBOUND TCP 192.168.2.10 192.168.4.20 3993 135 - - - - - - - -<br>2009-09-07 14:41:12 OPEN-INBOUND TCP 192.168.2.10 192.168.4.20 4002 4444 - - - - - - - -<br>2009-09-07 14:41:12 OPEN UDP 192.168.4.20 192.168.2.10 3027 69 - - - - - - - - | PERSONAL FIREWALL LOG → 135 OPEN-INBOUND TCP / 4444 OPEN-INBOUND TCP / 69 OPEN UDP / Source IP Address / Destination IP Address |
| | Security Log | Impact / Effect | **Event ID:**<br>592<br><br>**User ID:**<br>System<br><br>**Image File Name (IFN):**<br>• %WINDIR%\System32\tftp.exe<br>• %WINDIR%\System32\msblast.exe | 09/07/2009 14:41:12 Security Success Audit Detailed Tracking 592 NT AUTHORITY\SYSTEM SAHIB "A new process has been created:<br>New Process ID: 1016<br>Image File Name: C:\WINDOWS\system32\tftp.exe<br>Creator Process ID: 1228 User Name: SAHIB$ Domain: WORKGROUP<br>Logon ID: (0x0,0x3E7)<br><br>09/07/2009 14:41:25 Security Success Audit Detailed Tracking 592 NT AUTHORITY\SYSTEM SAHIB "A new process has been created:<br>New Process ID: 1752<br>Image File Name: C:\WINDOWS\system32\msblast.exe<br>Creator Process ID: 1228 User Name: SAHIB$ Domain: WORKGROUP<br>Logon ID: (0x0,0x3E7) | SECURITY LOG → ID: 592, User: System, IFN: %WINDIR%\System32\tftp.exe<br><br>ID: 592, User: System, IFN: %WINDIR%\System32\msblast.exe |
| | System Log | | **Event ID:**<br>7031<br>1074<br><br>**Image File Name (IFN):**<br>RPC service terminated unexpectedly<br>Windows Restart | 09/07/2009 14:41:29 Service Control Manager Error None 7031 N/A SAHIB **The Remote Procedure Call (RPC) service terminated unexpectedly.** It has done this 1 time(s). The following corrective action will be taken in 60000 milliseconds: Reboot the machine.<br>09/07/2009 14:41:29 USER32 Information None 1074 NT AUTHORITY\SYSTEM SAHIB The process winlogon.exe has initiated the restart of SAHIB for the following reason: No title for this reason could be found Minor Reason: 0xff<br>Shutdown Type: reboot Comment: **Windows must now restart** because the Remote Procedure Call (RPC) service terminated unexpectedly | SYSTEM LOG → ID: 7031 RPC service terminated unexpectedly<br><br>ID: 1074 Windows Restart |
| Network | Alert IDS Log | Activity | **Message:**<br>TFTP Get | [**] [1:1444:3] TFTP Get [**]<br>[Classification: Potentially Bad Traffic] [Priority: 2] | IDS ALERT LOG → TFTP Get |
| | | Alarm | **Source IP:**<br>SRC_IP: Victim<br><br>**Destination IP:**<br>Dest_IP: Attacker<br><br>**Destination Port:**<br>69 | 09/07-14:41:03.846382 192.168.4.20:3027 -> 192.168.2.10:69<br>UDP TTL:128 TOS:0x0 ID:337 IpLen:20 DgmLen:48 Len: 20 | IDS ALERT LOG → SRC_IP: Victim / Dest_IP: Attacker / Dest_Port: 69 |

Fig. 5 Summary of Blaster Traces on Victim's Log

In *Personal Firewall log*, ports 135 TCP, 4444 TCP and 69 UDP are considered as part of this victim's trace pattern due to the information gain from [7], [18] and [19]. These ports are known as vulnerable ports that can be used by malicious codes to exploit the system-level command shell on its victims. These ports also provide an inter-process communication mechanism that allows programs running on one host to execute code on remote hosts [20].

The source IP address of OPEN-INBOUND TCP indicates the remote host and the source IP address of OPEN UDP indicates the local host. While the destination IP address of OPEN-INBOUND TCP indicates the local host and the destination IP address of OPEN UDP indicates the remote host. The summarized of the IP address dependency is depicted in TABLE 1.

TABLE 1: IP ADDRESS DEPENDENCY

| Action | Source IP Address | Destination IP Address |
|---|---|---|
| OPEN-INBOUND | Remote Host | Local Host |
| OPEN | Local Host | Remote Host |

The patterns of the communication between Source IP address and Destination IP address indicate that the local host has permitted the TFTP service and the incoming traffic from the remote host. All communication activity is done by vulnerable ports open exploitation.

In *Security log*, the traces data from the security log shows that there is a new process created by system proves by the existence of *event id 592*. It has initiated the TFTP service used to download and upload the blaster worm code and execute the remote blaster worm code (msblast.exe). This activity is logged in *Image File Name* as *%WINDIR%\System32\tftp.exe* and *%WINDIR%\System32\msblast.exe.*

*System log* shows the traces data of the Blaster-infected machine stops its TFTP daemon after a transmission or after 20 seconds of TFTP inactivity by showing the new process created on event id 7031 and 1074 that indicates the RPC service terminated unexpectedly and windows restart respectively.

The *alert IDS log* shows that there is an activity called TFTP Get is occurred on port 69 UDP where the source IP

address from alert IDS log is similar as the destination IP address in the victim's firewall log and vice versa. These traces identify that there is a pattern exists on how the blaster worm initiates the client to download the worm code. The Source IP address is the victim and the Destination IP address is the attacker.

Action OPEN-INBOUND, OPEN-INBOUND and OPEN are known as a complete sequence of communication of blaster worm to gain access and upload the malicious codes to be exploited [21] where the OPEN action shows that an outbound session was opened to a remote host and the OPEN-INBOUND action shows that an inbound session was opened to the local host

*2) Attacker's Traces*

The attacker's data traces are extracted from the logs at the attacker's host and network log. The summary of the data traces on the attacker's and network logs are illustrated in Fig. 6 and the evidences are found in *personal firewall log, security log, system log* and *application log* and the details of the traces of the attacker's logs are discussed as following.

The traces data leaved in attacker's *Personal Firewall Log* shown the vulnerable ports that are used by the attacker to exploit the system-level command shell on its victims. The pattern of the traces data are OPEN TCP 135, OPEN TCP 4444 and OPEN-INBOUND UDP 69 as referred to [7], [18] and [19].

The source IP address of OPEN TCP indicates the local host and the destination IP address action indicates the remote host. While, the source IP address of OPEN-INBOUND UDP indicates the remote host and the destination IP address indicates the local host.

The patterns of source IP address and destination IP address indicate that the local host is opened an outbound session to the remote host which allow the local host transmit the payload (worm codes) to the remote host by exploiting the vulnerable ports open. The traces data from the *security log* shows that there is a new process created (Event ID: 592) that shows the blaster worm is activated based on the trace shows on the image file name.

In the *alert IDS log*, (Portscan) TCP Portsweep presents the pattern of scanning activity which shows the behavior of traditional worm attack in general and blaster worm attack in specific [22]. Therefore, this trace discovers that the owner of the source IP address is a potential attacker who launched the worm.

*3) Multi-step (Attacker/Victim) Trace Pattern*

Based on the extracted data from the logs at the victim's host and network log, the multi-step (Attacker/Victim)'s traces data is identified. The summary of the data traces on the multi-step and network logs are represented in Fig. 7 and the evidences are found in *personal firewall log, security log, system log* and *application log* and the details of the traces of the multi-step's logs are discussed.

There are two different patterns existing in *personal firewall log* that discover attacker and victim traces as shown in TABLE 2.

| Level | Evidence/Log | Category | Trace Pattern | Extracted Data | Conceptual Diagram |
|---|---|---|---|---|---|
| Host | Personal Firewall Log | Scan and Exploit | Action, Protocol and Destination Port: OPEN 135 TCP OPEN 4444 TCP OPEN-INBOUND 69 UDP<br><br>Source IP Address and Destination IP Address: Refer to Table 1 | 2009-09-07 14:41:09 **OPEN TCP** 192.168.2.10 192.168.4.20 3993 **135** - - - - - - - -<br>2009-09-07 14:41:11 **OPEN TCP** 192.168.2.10 192.168.4.20 4002 **4444** - - - - - - - -<br>2009-09-07 14:41:11 **OPEN-INBOUND UDP** 192.168.4.20 192.168.2.10 3027 **69** - - - -<br>- - - - | PERSONAL FIREWALL LOG → 135 OPEN TCP / 4444 OPEN TCP / 69 OPEN-INBOUND UDP / Source IP Address / Destination IP Address |
| | Security Log | Symptom | Event ID: 592<br><br>Image File Name (IFN): ~\blasterA.exe | 09/07/2009 14:40:01 Security Success Audit Detailed Tracking **592** Kamal TARMIZI "A new process has been created:<br>New Process ID: 1408<br>Image File Name: C:\Documents and Settings\Kamal\Desktop\**blasterA.exe**<br>Creator Process ID: 1492<br>User Name: Kamal<br>Domain: TARMIZI<br>Logon ID: (0x0,0x2273F) | SECURITY LOG → ID: 592 / User: Kamal / IFN: ~\blasterA.exe |
| Network | Alert IDS Log | Activity | Message: (Portscan) TCP Portsweep | [**] [122:3:0] **(portscan) TCP Portsweep** [**]<br>[Priority: 3] | IDS ALERT LOG → (Portscan) TCP Portsweep |
| | | Alarm | Source IP: SRC_IP: Attacker | 09/07-14:44:18.729996 **192.168.2.10** -> 192.168.11.1<br>PROTO:255 TTL:0 TOS:0x0 ID:3307 IpLen:20 DgmLen:166 | IDS ALERT LOG → SRC_IP: Attacker |

Fig. 6 Summary of Blaster Traces on Attacker's Log

| Level | Evidence/Log | Category | Trace Pattern | Extracted Data | Conceptual Diagram |
|---|---|---|---|---|---|
| Host | Personal Firewall Log | Scan and Exploit | **Destination Port:** 135, 4444, 69<br><br>**Action from Victim log:** OPEN-INBOUND, OPEN-INBOUND, OPEN<br><br>**Action from attacker log:** OPEN, OPEN, OPEN-INBOUND<br><br>**Protocol:** TCP, TCP, UDP<br><br>**Source IP Address and Destination IP Address:** Refer to Table 1 | 2009-09-07 14:41:09 OPEN-INBOUND TCP 192.168.2.10 192.168.4.20 3993 **135** - - - - -<br>2009-09-07 14:41:12 OPEN-INBOUND TCP 192.168.2.10 192.168.4.20 4002 **4444** - - - -<br>2009-09-07 14:41:12 OPEN UDP 192.168.4.20 192.168.2.10 3027 **69** - - - - - -<br><br>2009-09-07 14:45:24 OPEN TCP 192.168.4.20 192.168.11.20 4738 **135** - - - - - -<br>2009-09-07 14:45:27 OPEN TCP 192.168.4.20 192.168.11.20 4747 **4444** - - - - -<br>2009-09-07 14:45:27 OPEN-INBOUND UDP 192.168.11.20 192.168.4.20 3011 **69** - - - - - | PERSONAL FIREWALL LOG → **Victim**: 135 OPEN-INBOUND TCP, 4444 OPEN-INBOUND TCP, 69 OPEN UDP, Source IP Address, Destination IP Address<br><br>**Attacker**: 135 OPEN TCP, 4444 OPEN TCP, 69 OPEN-INBOUND UDP, Source IP Address, Destination IP Address |
| Host | Security Log | Impact / Effect | **Event ID:** 592<br><br>**Victim User ID:** System<br><br>**Attacker:** 5-21-725345543-1547161642-839522115-1003<br><br>**Image File Name (IFN):**<br>• %WINDIR%\System32 \tftp.exe<br>• %WINDIR%\System32 \msblast.exe | 09/07/2009  14:41:12 Security  Success Audit   Detailed Tracking **592** NT AUTHOR-ITY\SYSTEM   SAHIB5  "A new process has been created:<br>New Process ID: 1016<br>Image File Name:       C:\WINDOWS\system32\tftp.exe<br>Creator Process ID: 1228  User Name: SAHIB5 Domain: WORKGROUP<br>Logon ID: (0x0,0x3E7)<br><br>09/07/2009 14:41:25 Security  Success Audit Detailed Tracking **592** NT AUTHOR-ITY\SYSTEM   SAHIB  "A new process has been created:<br>New Process ID: 1752<br>Image File Name:       C:\WINDOWS\system32\msblast.exe<br>Creator Process ID: 1228    User Name: SAHIB5 Domain: WORKGROUP<br>Logon ID: (0x0,0x3E7)<br><br>09/07/2009    14:42:40 Security Success Audit   Detailed Tracking **592 S-1-5-21-725345543-1547161642-839522115-1003**   SAHIB  "A new process has been created:<br>New Process ID: 288<br>Image File Name:       C:\WINDOWS\system32\msblast.exe<br>Creator Process ID:   1748   User Name:Shahrin Domain:    SAHIB<br>Logon ID :(0x0,0xD8B5) | SECURITY LOG → **Victim**: ID: 592, User: System, IFN: %WINDIR%\System32 \tftp.exe<br><br>ID: 592, User: System, IFN: %WINDIR%\System32 \msblast.exe<br><br>**Attacker**: ID: 592, User: S-1-5-21-725345543-1547161642-839522115-1003, IFN: %WINDIR%\System32\msblast.exe |
| Host | System Log | | **Event ID:** 7031, 1074<br><br>**Image File Name (IFN):**<br>• RPC service terminated unexpectedly<br>• Windows Restart | 09/07/2009     14:41:29      Service Control Manager  Error None **7031** N/A SAHIB<br>**The Remote Procedure Call (RPC) service terminated unexpectedly.** It has done this 1 time(s). The following corrective action will be taken in 60000 milli-seconds: Reboot the machine.<br><br>09/07/2009     14:41:29      USER32 Information None  **1074**<br>NT AUTHORITY\SYSTEM  SAHIB The process winlogon.exe has initiated the restart of SAHIB for the following reason: No title for this reason could be found<br>Minor Reason: 0xff<br> Shutdown Type: reboot  Comment: **Windows must now restart** because the Remote Procedure Call (RPC)  service terminated unexpectedly | SYSTEM LOG → **Victim**: ID: 7031, RPC service terminated unexpectedly<br><br>ID: 1074, Windows Restart |
| Network | Alert IDS Log | Activity | **Victim Message:** TFTP Get<br><br>**Attacker Message:** (Portscan) TCP Portsweep | [**] [1:1444:3] **TFTP Get** [**]<br>[Classification: Potentially Bad Traffic] [Priority: 2]<br><br>[**] [122:3:0] **(portscan) TCP Portsweep** [**]<br>[Priority: 3] | IDS ALERT LOG → **Victim**: TFTP Get<br><br>**Attacker**: (Portscan) TCP Portsweep |
| Network | Alert IDS Log | Alarm | **Victim**<br>Source IP: SRC_IP: Victim<br><br>Destination IP: Dest_IP: Attacker<br><br>Destination Port: 69<br><br>**Attacker**<br>Source IP: SRC_IP: Attacker | 09/07-14:41:03.846382 **192.168.4.20:3027 -> 192.168.2.10:69**<br>UDP TTL:128 TOS:0x0 ID:337 IpLen:20 DgmLen:48 Len: 20<br><br>09/07-14:45:38.384318 **192.168.4.20** -> 192.168.11.1<br>PROTO: 255 TTL: 0 TOS: 0x0 ID: 3395 IpLen: 20 DgmLen: 164 | IDS ALERT LOG → **Victim**: SRC_IP: Victim, Dest_IP: Attacker, Dest_Port: 69<br><br>**Attacker**: SRC_IP: Attacker |

Fig. 7: Blaster Traces on Multi-step (Victim/Attacker) Log

TABLE 2: TRACES DATA ON PERSONAL FIREWALL LOG

| Victim | Attacker |
|---|---|
| **Action, Protocol and Destination port:**<br><br>OPEN-INBOUND TCP 135, OPEN-INBOUND TCP 4444 and OPEN UDP 69 - These ports are known as vulnerable ports that provide hole and opportunities to the attacker to exploit in order to gain access to the system. | **Action, Protocol and Destination port:**<br><br>OPEN TCP 135, OPEN TCP 4444 and OPEN -INBOUND UDP 69 - These ports are known as vulnerable ports that can be used by the attacker to exploit the system-level command shell on its victims. |
| **Source IP Address and Destination IP Address:**<br><br>OPEN-INBOUND TCP - the source IP address indicates the remote host and the destination IP address indicates the local host. OPEN UDP - the source IP address indicates the local host and the destination IP address indicates the remote host. | **Source IP Address and Destination IP Address:**<br><br>OPEN TCP - the source IP address indicates the local host and the destination IP address indicates the remote host. OPEN-INBOUND UDP - the source IP address indicates the remote host and the destination IP address indicates the local host. |

From the victim perspective *(OPEN-INBOUND TCP 135, OPEN-INBOUND TCP 4444 and OPEN UDP 69)*, the patterns shows that the source IP address and destination IP address indicate that the local host is permitted the *TFTP* service and the incoming traffic from the remote host.

While, from the attacker perspective *(OPEN TCP 135, OPEN TCP 4444* and *OPEN -INBOUND UDP 69)*, the patterns of source IP address and destination IP address indicate that the local host is opened an outbound session to the remote host which allow the local host transmit the payload (worm codes) to the remote host.

All communication activity is done by vulnerable ports open exploitation. Therefore, the traces data found are significant to the multi-step attack (victim/attacker) where this host was infected (act as victim) and as long as the computer was infected with the worm code *(msblast)*, it (act as attacker) continued to generate traffic to attempt to infect other vulnerable computers [23].

The traces data in TABLE 3 are taken from the *security log* in Fig. 9 and it shows that there is a new process created by system which initiates the *TFTP* service. This service is used to receive and sent the blaster worm code and execute the blaster worm code (*msblast.exe*) remotely.

TABLE 3: TRACES DATA ON SECURITY LOG

| Victim | Attacker |
|---|---|
| **Event ID:** 592 | **Event ID:** 592 |
| **User ID:** System | **User ID:** S-1-5-21-725345543-1547161642-839522115-1003 |
| **Image File Name:** %WINDIR%\System32\tftp.exe, %WINDIR%\System32\msblast.exe | **Image File Name** %WINDIR%\System32\msblast.exe |

The pattern in TABLE 3 has indicates that this host is a victim of blaster worm attack. Another traces data found is a new process created (Event ID: 592) by unidentified user (S-1-5-21-725345543-1547161642-839522115-1003) which executed the *msblast* as shown on the image file name. It identify that this host was infected previously and automatically attempt to transfer the worm code by generating traffic to exploit other vulnerable computers.

*System log* shows the traces data of the Blaster-infected machine stops by showing the new process created on event id 7031 and 1074 that indicates the RPC service terminated unexpectedly and windows restart respectively. This pattern is significant with the victim pattern in which if the host is infected by blaster worm, the RPC service is terminated unexpectedly by exploiting the RPC DCOM and force the windows restart.

As depicted in TABLE 4, the *alert IDS log* shows that there are traces found on TFTP Get and (Portscan) TCP Portsweep activities for victim and attacker respectively.

The *TFTP Get* activity on port *69* trace indicates that there is a pattern exists on how the blaster worm initiates the client to download the worm code. The source IP address is the victim and the destination IP address is the attacker. On the other hand, the *(Portscan) TCP Portsweep* trace proves that there is scanning activity on the vulnerable open port. This trace indicates that the source IP address is the attacker who activated the worm. Both traces found in TABLE 4 are significant to the pattern that found in victim and attacker as depicted in Fig. 8 and Fig. 9 respectively.

TABLE 4: TRACES DATA ON ALERT IDS LOG

| Victim | Attacker |
|---|---|
| **Message:** TFTP Get | **Message:** (Portscan) TCP Portsweep |
| **Source IP Address and Destination IP Address:** Source IP address indicates the victim and the Destination IP address indicates the attacker | **Source IP Address:** Source IP address indicates the attacker |
| **Destination Port:** 69 | |

Based on the analysis, the researchers have identified the significant attributes from the victim, attacker and multi-step traces data. These findings are further use to construct the proposed worm trace pattern.

V. PROPOSED WORM TRACE PATTERN

This research proposed the worm trace pattern based on victim, attacker and multi-step point of view. The following section describes the details.

*A. Victim's Trace Pattern*

Victim's trace pattern is useful for forensic in order to provide clear view on how the victim attacked by the potential attacker. According to the analysis and findings from Fig. 5, the overall Blaster victim's trace pattern is summarized in Fig.8.

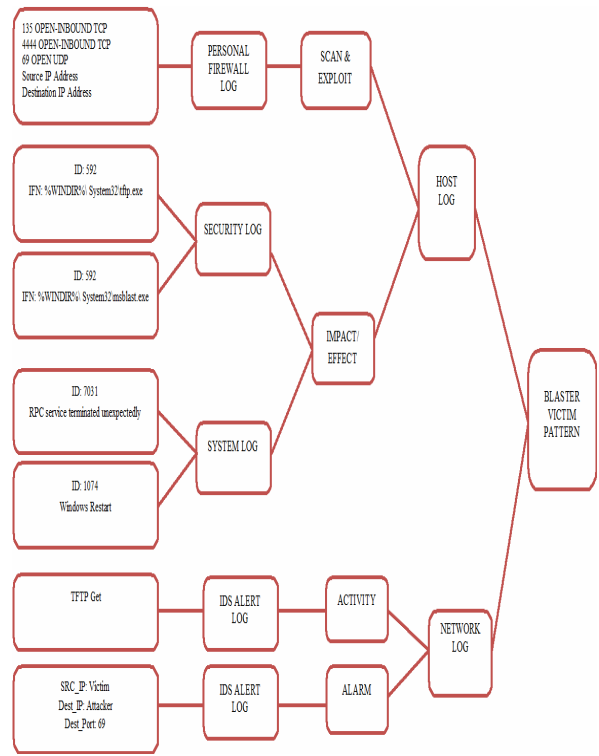

Fig.8 Proposed Blaster Victim's Trace Pattern

In Fig. 8, the traces data indicated the blaster worm pattern at the victim's host used port 135 TCP to permit the scanning and transmitting RPC DCOM exploit codes from remote host which launch the windows shell to initiate worm code download used port 4444 TCP.

Then it launched the TFTP client service using port 69 for downloading the worm code. The traces on TFTP Get on port 69 UDP also found in the network log that supports all the traces found on the host log.

*B. Attacker's Trace Pattern*

Attacker's trace pattern provides a systematic description of the attack goals and attack strategies for defending against and tracing the attack. This pattern is useful to guide forensic investigators in searching the evidence and provide a structured method for obtaining and representing relevant network forensic information.

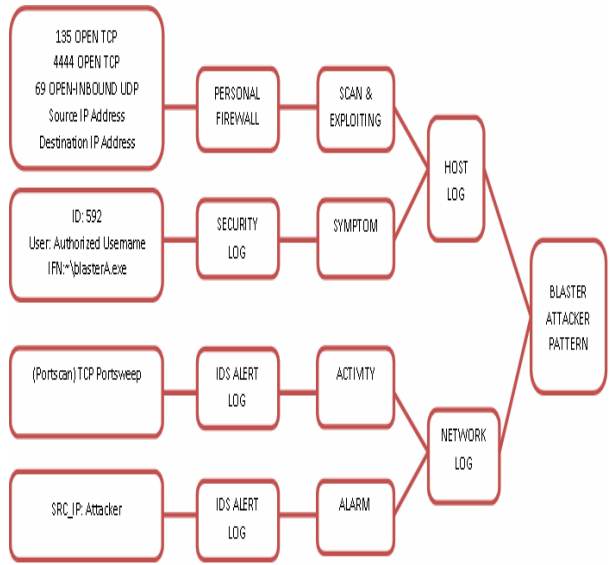

Fig. 9  Proposed Blaster Attacker's Trace Pattern

The overall Blaster attacker's trace pattern depicted in Fig. 9 indicate the blaster worm pattern at the attacker's host used port *135 TCP* to allow the local host scan and transmit *RPC DCOM* exploit codes to the remote host which launch the windows shell to initiate worm code download used port *4444 TCP*. Then it launched the *TFTP* client service using port *69* to permit the client (remote host) download the worm code from the local host. The activity of *TCP Portsweep* trace also existed in the network log that supports all the traces found on the host log.

### C. Multi-step (Attacker/Victim) Trace Pattern

Multi-step's trace pattern is used as a guide for forensic investigators to reveal and prove the true attacker or victim. This trace pattern is a combination of victim's and attacker's trace pattern in which the traces data is extracted from a log for the same host.

The traces data on multi-step at the host's logs from victim/attacker perspective illustrated in Fig. 10 indicate that the blaster worm used port 135 TCP to permit the scanning activity and it is supported by the traces found in network logs that show *(Portscan) TCP Portsweep* activities.

The worm then transmit RPC DCOM exploit codes from remote host which launch the windows shell to initiate downloading the worm code using port 4444 TCP and it launched the TFTP client service on port 69. This worm activity is shown by the traces found in network logs that confirm the existence of *TFTP Get on port 69 UDP* activities. Once the host is infected (act as victim), it's (act as attacker) then generate traffic; attempt to infect other vulnerable hosts.

The source IP address from host log indicates that the remote host is the victim and the destination IP address which is the local host is the attacker. Hence, multi-step (victim/attacker) trace pattern could identify the true victim or attacker.

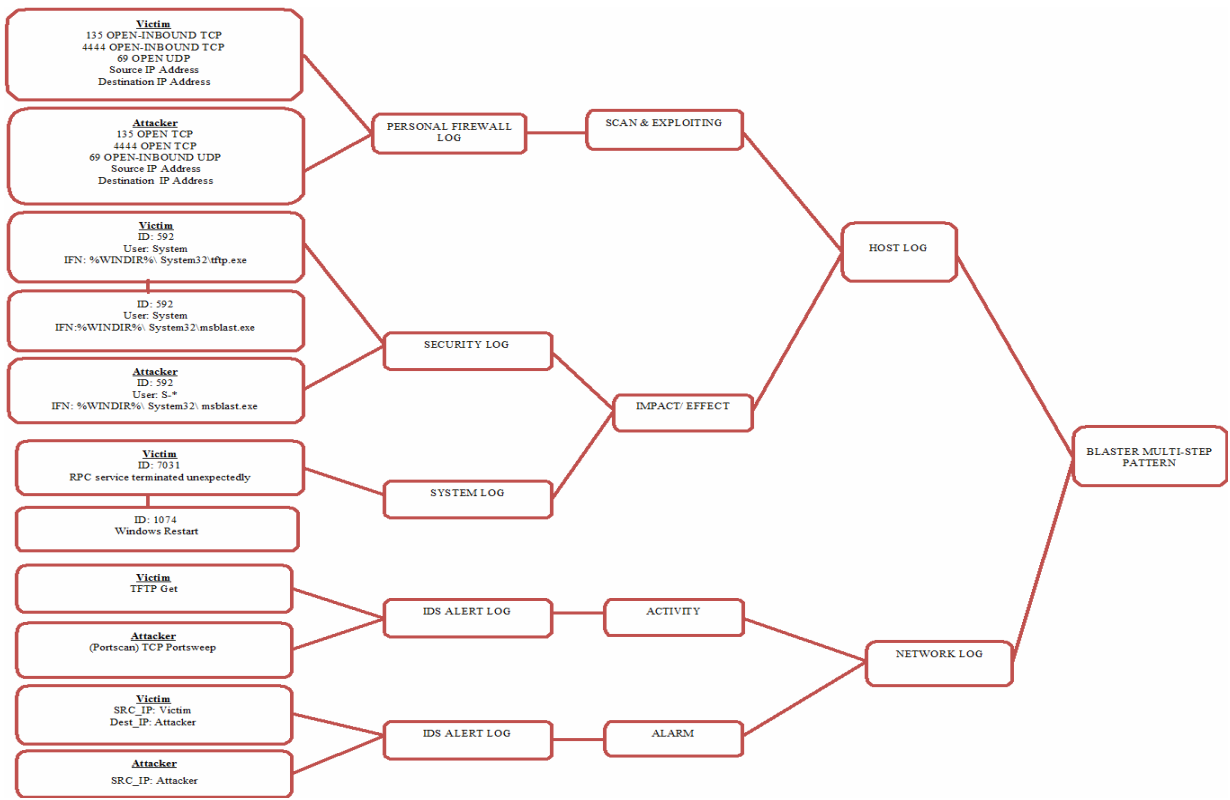

Fig. 10 Proposed Multi-step (Victim/Attacker) Trace Pattern

## VI. CONCLUSIONS AND FUTURE WORKS

Trace pattern of an attack in an attack scenario is constructed by analyzing various logs from heterogeneous devices in victim, attacker and victim/attacker (multi-step) perspectives. These trace patterns offer a systematic description of the impact of the attack, the attack goals and attack steps along with strategies for defending against and tracing the attack. For example, personal firewall logs provide information on how the attacker entered the network and how the exploits were performed; meanwhile event logging such as security log, system log and application log enables network administrators to collect important information such as date, time and result of each action during the setup and execution of an attack. Therefore, the propose victim, attacker and multi-step (victim/attacker) trace patterns in this paper can be extended into research areas in alert correlation and computer forensic investigation.